\documentclass[usenatbib,letterpaper]{mn2e}

\usepackage{amsmath,amssymb, graphics}
\usepackage{pgf,tikz}
\usepackage{bm}
\usepackage{graphicx}
\usepackage{epstopdf}
\usepackage{float}

\newcommand{\Mp}{M_\text{p}}
\newcommand{\Rp}{R_\text{p}}
\newcommand{\Ms}{M_*}
\newcommand{\bTst}{{\bm T}_*}

\newcommand{\omst}{\omega_*}
\newcommand{\pomst}{\bar 
\omega_*}
\newcommand{\bTsp}{{\bm T}_\text{sp}}

\newcommand{\pomsp}{\bar \omega_\text{sp}}
\newcommand{\bTsg}{{\bm T}_\text{sg}}
\newcommand{\bl}{\bm {\hat l}}
\newcommand{\blp}{\bm {\hat l}_\text{p}}
\newcommand{\bs}{\bm {\hat s}_\text{p}}
\newcommand{\rin}{{r_\text{in}}}
\newcommand{\rout}{{r_\text{out}}}
\newcommand{\Md}{M_\text{d}}

\newcommand{\der}{\text{d}}
\newcommand{\pd}{\partial}
\newcommand{\Om}{\Omega}
\newcommand{\om}{\omega}
\newcommand{\bgp}{\beta_\text{p}}

\newcommand{\bj}{\bm j}

\newcommand{\bn}{\bm{\hat n}}

\newcommand{\ag}{\alpha}
\newcommand{\bg}{\beta}

\newcommand{\Sg}{\Sigma}
\newcommand{\Sgout}{\Sigma_\text{out}}
\newcommand{\sg}{\sigma}

\newcommand{\rsp}{r_\text{sp}}
\newcommand{\rst}{r_*}

\newcommand{\one}{{(1)}}

\newcommand{\btimes}{{\bm \times}}
\newcommand{\bcdot}{{\bm \cdot}}

\newcommand{\be}{\begin{equation}}
\newcommand{\ee}{\end{equation}}

\begin{document}


\title[Extended Transiting Disks and Rings]{Extended Transiting Disks and Rings Around \\
 Planets and Brown Dwarfs: Theoretical Constraints}


\author[J. J. Zanazzi and Dong Lai]{J. J. Zanazzi$^{1}$\thanks{Email: jjz54@cornell.edu}, and Dong Lai$^{1}$ \\
$^{1}$Cornell Center for Astrophysics, Planetary Science, Department of Astronomy, Cornell University, Ithaca, NY 14853, USA}




\maketitle
\begin{abstract}
Newly formed planets (or brown dwarfs) may possess disks or
rings that occupy an appreciable fraction of the planet's Hill
sphere and extend beyond the Laplace radius, where the tidal torque from the host star dominates over the torque from the oblate planet. Such a disk/ring can exhibit unique, detectable transit
signatures, provided that the disk/ring is significantly misaligned
with the orbital plane of the planet. There exists tentative evidence
for an extended ring system around the young K5 star 1 SWASP J140747-354542.
We present a general theoretical study of the inclination (warp)
profile of circumplanetary disks under the combined influences of the
tidal torque from the central star, the torque from the oblate planet
and the self-gravity of the disk.  We calculate the 
equilibrium
warp profile (``generalized Laplace surface'') and investigate the
condition for coherent precession of the disk.  We find that to
maintain non-negligible misalignment between the extended outer disk and the
planet's orbital plane, and to ensure coherent disk
precession, the disk surface density must be sufficiently
large so that the self-gravity torque overcomes the tidal torque from
the central star. Our analysis and quantitative results can be used to constrain the parameters
of transiting circumplanetary disks that may be detected in the future.
\end{abstract}





\section{Introduction}
\label{sec:Intro}

The age is nearing when direct observations of circumplanetary disks and rings become a reality through photometry.  A number of studies have investigated the detectability and observational signatures of circumplanetary disks/rings \citep{BarnesFourtney(2004), Ohta(2009),SchlichtingChang(2011),TusnskiValio(2011),Zuluaga(2015)}.  Although observational searches for exo-rings have been carried out, most are inconclusive \citep{Brown(2001),Heising(2015),Santos(2015)}.  These searches focused on hot Jupiters, which have Hill radii $r_H \equiv a (\Mp/3\Ms)^{1/3}$ (where $a$ is the planetary semi-major axis, $\Mp$ is the planet mass, and $\Ms$ is the mass of the host star) comparable to their planetary radii $\Rp$.  For this reason, these circumplanetary disks could not have outer radii $\rout$ significantly larger than their respective planetary radii.

\cite{Mamajek(2012)} discovered that the light curve of a young ($\sim$ 16 Myr) K5 star 1 SWASP J140747-354542 (hereafter J1407) exhibited a complex series of eclipses that lasted 56 days around the month of April 2007.  The central deep ($>3$ mag) eclipse was surrounded by two pairs of 1 mag eclipses occuring at $\pm 12$ and $\pm 26$ days.  They proposed that these eclipses where caused by a large ring system orbiting an unseen companion J1407b (see also \citealt{vanWerkhoven(2014)}).  Other explanations were considered but deemed unlikely.  Follow-up observations by \cite{Kenworthy(2015)} constrain the companion mass to $<80 \,M_\text{J}$ (where $M_\text{J}$ is the mass of Jupiter) and semi-major axis (for circular orbits) to $a \simeq 2.2 - 5.6 \, \text{AU}$ ($3\sg$ significance).  Thus, J1407b is most likely a giant planet or brown dwarf in a 3.5-14 year orbit around the primary star.  Modeling the eclipse light curve with a series of inclined, circular optically thick rings gave a best fit outer disk radius of $\sim 0.6 \, \text{AU}$, a significant fraction of the companion's Hill radius \citep{vanWerkhoven(2014),KenworthyMamajek(2015)}.  The disk/ring system also contains gaps, which may be cleared by exomoons orbiting around J1407b.


Currently, the ring/disk interpretation of the J1407 light curve
remains uncertain, although no serious alternatives have been explored in detail. The ring/disk interpretation can be tested in the coming
years if another eclipse event is detected, while a non-detection
would put the model under increasing strain. In any case, the possible
existence of such a ring system naturally raises questions about the
formation of inclined, extended disks/rings around giant planets and brown
dwarfs. In order to produce a transiting signature, the disk/ring must be 
inclined with respect to the orbital plane.
How are such inclinations produced and maintained?

For giant planets, the inclination of the disk/ring may be tied
to the obliquity of the planet due to its rotation-induced quadrupole.
The obliquity may be excited through secular spin-orbit resonances, as
in the case of Saturn \citep{HamiltonWard(2004),WardHamilton(2004),VokrouhlickNesvorn(2015)},
or through impacts with planetesimals of sufficiently large masses \citep{LissauerSafronov(1991)}.
In the case of brown dwarfs, which are thought to form
independently of the primary, the disk could be ``primordially'' 
misaligned with respect to the binary orbit because of the turbulent motion 
of gas in the star forming environment \citep{Bate(2009),Bate(2010),Tokuda(2014)}.

In this paper, we will address the following question: Under what conditions can
an extended disk/ring around a planet or brown dwarf maintain its
inclination with respect to the planet's orbit in order to exhibit a
transit signature? As discussed in Section 2, even when the disk is
safely confined within the planet's Hill sphere, the outer region of
the disk can still suffer significant tidal torque from the host star.  This
tidal torque tends to induce differential precession of the disk.
Without any internal forces, the disk will lose coherence in shape and
inclination. In the presence of dissipation, the disk may reach a 
equilibrium warp profile
(called ``Laplace surface'') 
in which the outer region of the disk [beyond the Laplace radius; see
Eq.~\eqref{eq:rL} below] becomes aligned with the orbital plane.

In gaseous disks, hydrodynamic forces work to keep the disk
coherent, through bending waves \citep{IvanovIllarionov(1997),PapaloizouLin(1995),LubowOgilvie(2000)}
or viscosity \citep{PapaloizouPringle(1983),Ogilvie(1999)}. But the
rapid variability in the photometric data for the inferred ring system
around J1407b implies that the disk/ring system is quite thin, with
a ratio of the scaleheight to radius of order $H/r \sim 10^{-3}$
\citep{vanWerkhoven(2014)}, with significant gaps in the disk \citep{Mamajek(2012),KenworthyMamajek(2015)}. 

It is unlikely that hydrodynamical forces are sufficiently strong 
to maintain the disk's coherence (see Section 5.2).


Another plausible internal torque is self-gravity (e.g., Ward 1981;
Touma et al.~2009; Ulubay-Siddiki et al~2009). This is the possibility we
will focus on in this paper. Of particular relevance is the work by 
Ward (1981), who studied the warping of a massive self-gravitating disk in an
attempt to explain the inclination of Iapetus, Saturn's moon, with
respect to the local Laplace surface. He found that self-gravity
of the circumplanetary disk which formed Saturn's satellites could
significantly modify the equilibrium inclination/warp profile.  

In this paper, we re-examine the warp dynamics of self-gravitating
circumplanetary disks in light of the possible detection extended transiting disks.
We consider general (possibly large) planetary obliquities, and study
both equilibrium disk warp and its time evolution. 
Our goal is to derive the conditions (in terms of disk mass and density profile)
under which an extended circumplanetary disk/ring maintain its inclination 
with respect to the planet's orbit.
In Section~\ref{sec:SS+SG}, we study the equilibrium inclination/warp
profile of the disk, which we will call the \textit{Generalized Laplace Surface},
under the influences of torques from the
oblate planet, the distant host star, and disk self-gravity.
We show that if the disk is sufficiently massive, the outer region of the disk 
can maintain significant inclination relative to the planet's orbit.
In Section 4, we study the time evolution of disk warp, 
including the stability of the generalized Laplace surfaces, and the condition 
required for the disk to be capable of precessing coherently.  
We summarize our results and discuss their implications in Section~\ref{sec:conc}

Although it is unknown if the object J1407b is a planet or brown
dwarf, we will refer to J1407b as a ``planet" throughout the rest of
the paper.

\section{External Torques and the Laplace Surface}
\label{sec:System+Torques}

Consider a planet (mass $\Mp$) in a circular orbit around a central star (mass $\Ms$) with orbital semi-major axis $a$.  We denote the unit orbital angular momentum vector by $\blp$.     We take the circumplanetary disk to extend from radius  $r = \rin$ to $r = \rout$, as measured from the center of the planet.  We assume that the disk is circular.  In general, the angular momentum unit vector at each annulus of the disk is a function of radus and time, and is specified by $\bl = \bl(r,t)$.

The circumplanetary disk experiences two external torques, from the host star and from the planet's quadrupole.  Averaging over the orbit of the planet, to leading order in the ratio $r/a$, the tidal torque per unit mass from the star exerted on a disk annulus with unit angular momentum $\bl$ is 
\be
\label{eq:Tst}
\bTst = \frac{3 G \Ms r^2}{4 a^3} \big( \bl \bcdot \blp \big) \big( \bl \btimes \blp \big).
\ee
The quadrupole moment of the planet is related to its $J_2$ parameter by $I_3 - I_1 =J_2 \Mp \Rp^2$, where $\Rp$ is the radius of the planet,
and $J_2$ depends on the planet's rotation rate $\Om_\text{p}$ via $J_2 ~= ~(k_2/3)(\Om_\text{p}^2 \Rp^3/G \Mp).$
  The Love number $k_2$ is of order $0.4$ for giant planets.  The torque from the spinning planet on the disk annulus is
\be
\label{eq:Tsp}
\bTsp = \frac{3 G \Mp \Rp^2 J_2}{2 r^3} \big(\bl \bcdot \bs \big) \big( \bl \btimes \bs \big),
\ee
where $\bs$ is the unit vector along the planet's spin axis.

In general, when $\bl$, $\blp$, and $\bs$ are not parallel to each other, $|\bTst|$ dominates at large $r$ while $|\bTsp|$ dominates at small $r$.  The radius where $|\bTst| \sim |\bTsp|$ defines the Laplace radius
\be
\label{eq:rL}
r_L \equiv \left(2 J_2 \frac{\Mp}{\Ms} \Rp^2 a^3 \right)^{1/5} = \left( 6 J_2 \Rp^2 r_H^3 \right)^{1/5},
\ee
where $r_H \equiv a (\Mp/3\Ms)^{1/3}$ is the Hill radius \citep{Tremaine(2009)}.  Tidal truncation and dynamical stability require that the outer radius of the disk be less than a fraction of $r_H$, i.e. $\xi \equiv \rout/r_H \lesssim 0.4$ (e.g. \citealt{QuillenTrilling(1998),AyliffeBate(2009),MartinLubow(2011),LehebelTizcareno(2015)}).  Thus the ratio of $r_L$ to $\rout$ is given by
\begin{align}
\frac{r_L}{\rout} = &\left( \frac{6 J_2 \Rp^2}{\xi^3 r_\text{out}^2} \right)^{1/5}
\nonumber \\
= &0.18 \left( \frac{J_2}{10^{-2}} \right)^{1/5} \left( \frac{\rout}{0.2 r_H} \right)^{-3/5} 
\nonumber \\
&\times \left( \frac{\Rp}{R_\text{Jup}} \right)^{2/5} \left( \frac{\rout}{0.1 \, \text{AU}} \right)^{-2/5}.
\end{align}
where we have scaled $J_2$ to the value appropriate to gas giants in our Solar System, and $\rout$ appropriate to the claimed ring system in J1407 \citep{vanWerkhoven(2014)}.

In the presence of dissipation in the disk, we may expect $\bl(r,t)$ to evolve toward
the equilibrium state, in which 
\be
\bTst + \bTsp = 0.
\label{eq:steady}
\ee
The equilibrium orientation of the disk $\bl(r)$, which defines the
Lapace surface \citep{Laplace(1805),Tremaine(2009)}, lies in the plane
spanned by the vectors $\bs$ and $\blp$.  Throughout this
paper, we assume that the planet's spin angular momentum is much larger
than the disk angular momentum, so that $\bs$ is fixed in time.
Let $\bgp$ be the planetary obliquity (the angle between $\bs$ and $\blp$)
and $\bg(r)$ be the warp angle of the disk [the angle between $\bl(r)$ and
$\blp$]. Equation \eqref{eq:steady} may be reduced to 
\begin{align}
\label{eq:LSurf}
0 = &z^2 \cos\bg(z) \sin \bg(z) 
\nonumber\\
&+ \frac{z_L^5}{z^3} \cos \big[\bg(z) - \bgp \big] \sin \big[ \bg(z) - \bgp \big],
\end{align}
where we have defined the dimensionless Laplace radius $z_L$ and
radial coordinate $z$ by 
\be
z_L \equiv r_L/\rout,\quad
z \equiv r/\rout.
\ee

\begin{figure}
\centering
\includegraphics[scale=0.42]{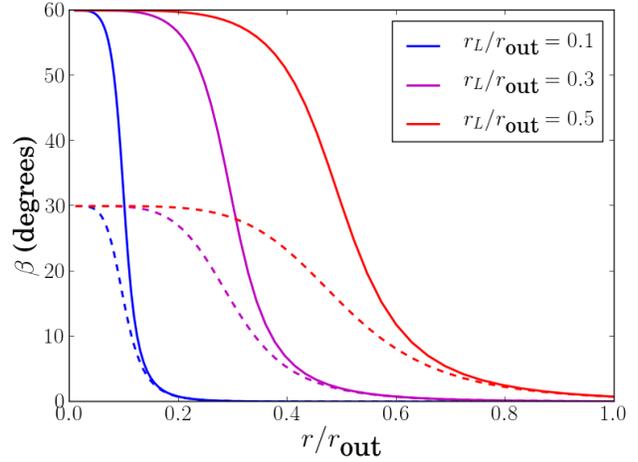}
\caption{
Equilibrium disk inclination profile (Laplace surface without self-gravity).  The quantity $\bg$ is the angle between $\bl$ and $\blp$.  The different lines are for $r_L/\rout = 0.1$ (blue), $0.3$ (magenta) and $0.5$ (red).  The planetary obliquity $\bgp$ is assumed to be $ 60^\circ$ (solid lines) and $30^\circ$ (dotted lines).}
\label{fig:laplace}
\end{figure}

Figure~\ref{fig:laplace} depicts the solutions to Eq.~\eqref{eq:LSurf} for
$\bgp = 30^\circ, 60^\circ$ and $z_L = 0.2, 0.5$.  Clearly, 
in the absence of any internal torque, the outer region of the disk (beyond $\sim 2 r_L$)
is highly aligned with the planetary orbit, with 
\be
\label{eq:betalap}
\beta (r)\simeq \left({r_L\over r}\right)^5\cos \bgp \sin \bgp.
\ee
Such an aligned outer disk would not produce the transit signal
claimed in the J1407 system. To maintain significant inclination in
the outer disk, some internal torques are needed. We consider the
effect of self-gravity in the next section.


\section{Generalized Laplace Surface: Equilibrium with Self-Gravity}
\label{sec:SS+SG}

In this section, we consider the influence of self-gravity on the 
equilibrium warp profile $\bl(r)$ of the disk.  Let the surface density of the disk be $\Sg = \Sg(r)$. The torque acting on the disk due to its own self-gravity is 
approximately given by 
\begin{align}
\label{eq:Tsg}
\bTsg \simeq \frac{\pi G}{2} \int_\rin^\rout \der r' & \frac{r' \Sg(r')}{\max(r,r')} \chi b_{3/2}^\one(\chi) 
\nonumber \\
&\times \big[ \bl(r) \bcdot \bl(r') \big] \big[ \bl(r) \btimes \bl(r') \big],
\end{align}
where $\chi = \min(r,r')/\max(r,r')$ and $b_{3/2}^\one(\chi)$ is the Laplace coefficient
\be
b_{3/2}^\one(\chi) = \frac{2}{\pi} \int_0^\pi \frac{\cos\theta \der \theta}{(1 - 2\chi \cos\theta + \chi^2)^{3/2}}.
\ee
Eq.~\eqref{eq:Tsg} is an approximation which recovers two limits:  When $|\bl(r) \times \bl(r')| \ll 1$, it reduces to Eq. (8) of \cite{Tremaine(1991)} and Eq. (47) of \cite{TremaineDavis(2014)}; when $\chi \ll 1$, $b_{3/2}^\one(\chi) \simeq 3 \chi$ \citep{MurrayDermott(1999)} and we recover the quadrupole approximation:
\begin{align}
\label{eq:sgquad}
\bTsg \simeq \frac{3\pi G}{2} \int_\rin^\rout  \der r' & \frac{r' \Sg(r')}{\max(r,r')}\chi^2 
\nonumber \\
&\times \big[ \bl(r) \bcdot \bl(r') \big] \big[ \bl(r) \btimes \bl(r') \big].
\end{align}


The integrand of Eq.~\eqref{eq:Tsg} becomes invalid when $\chi \sim 1$
and $|\bl(r) \btimes \bl(r')| \sim 1$ (i.e., when two close-by annuli 
have a large mutual inclination), and a different
formalism is needed to calculate the torque acting on a disk from its
own self-gravity
(e.g. \citealt{Kuijken(1991),ArnaboldiSparke(1994),Ulubay-Siddiki(2009)}).
In the appendix, we review the exact equations for calculating 
internal self-gravity torques for arbitrary $\chi$ and
$|\bl(r)\btimes \bl(r')|$.  Our numerical calculations based on 
these exact (but much more complicated) equations show that they provide 
only minor quantitative corrections to
the disk warp profile and the inclination at the outer disk radius.
For this reason, we will use the much simpler approximation \eqref{eq:Tsg} for the 
remainder of this paper.

For concreteness, we consider a power-law surface density profile
\be
\label{eq:Sg}
\Sg(r) = \Sgout \left( \frac{\rout}{r} \right)^p.
\ee
Then the disk mass is (assuming $\rin \ll \rout$)
\be
\Md \simeq \frac{2\pi}{2-p} \Sgout r_\text{out}^2,
\ee
and the total disk angular momentum is
\be
L_\text{d} \simeq \frac{4\pi}{5-2p} \Sgout r_\text{out}^2 \sqrt{G \Mp \rout}.
\ee

\begin{figure}
\centering
\includegraphics[scale=0.42]{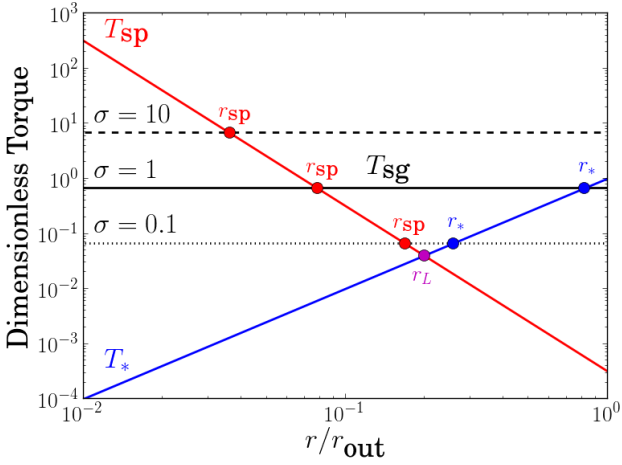}
\caption{Torques on the disk based on the estimates \eqref{eq:sgest}-\eqref{eq:spest} and normalized by $|\bTst(\rout)|$.  The tidal torque from the star $(\bTst)$ is in blue, and the torque from the spinning planet $(\bTsp)$ is in red.  The torque from self-gravity $(\bTsg)$ is in black, with three values of $\sg$ [see Eq.~\eqref{eq:sigma}] as indicated, all for $p=1$ [see Eq.~\eqref{eq:Sg}].  The three critical radii in the disk $(\rsp,r_L,\rst)$ are marked.}
\label{fig:torques}
\end{figure}

It is useful to compare the magnitude of $|\bTsg|$ to the external torques acting on the disk (see Fig.~\ref{fig:torques}).  Ignoring geometrical factors relating to the angles between $\bl, \blp$ and $\bs$, we have to an order of magnitude [see Eqs. \eqref{eq:Tsg}, \eqref{eq:Tst}, and \eqref{eq:Tsp}]
\begin{align}
\label{eq:sgest}
|\bTsg| &\sim \pi G \Sg(r) r \\
\label{eq:stest}
|\bTst| &\sim \frac{3 G \Ms r^2}{4 a^3} \\
\label{eq:spest}
|\bTsp| &\sim \frac{3 G \Mp \Rp^2 J_2}{2 r^3}.
\end{align}
Thus
\begin{align}
\label{eq:rst}
\frac{|\bTsg|}{|\bTst|} &\sim \frac{2(2-p)}{3} \sg \left( \frac{\rout}{r} \right)^{1+p} \equiv \left( \frac{\rst}{r} \right)^{1+p}, \\
\label{eq:rsp}
\frac{|\bTsg|}{|\bTsp|} &\sim \frac{2(2-p)}{3} \frac{\sg}{z_L^5} \left( \frac{r}{\rout} \right)^{4-p} \equiv \left( \frac{r}{\rsp} \right)^{4-p},
\end{align}
where we have defined the dimensionless parameter $\sg$ (which measures $|\bTsg|/|\bTst|$ at $r = \rout$) as
\be
\label{eq:sigma}
\sg \equiv \frac{\Md}{\Ms} \left( \frac{a}{\rout} \right)^3 = 0.38 \left( \frac{\rout}{0.2 \, r_H} \right)^{-3} \left( \frac{\Md}{10^{-3} \Mp} \right).
\ee
In Eqs. \eqref{eq:rst} and \eqref{eq:rsp}, $\rst$ and $\rsp$ are set by $|\bTsg|/|\bTst|\sim1$ and $|\bTsg|/|\bTsp| \sim1$ respectively.  Recall the Laplace radius $r_L$ is set by $|\bTst|\sim|\bTsp|$.  For radii $r\lesssim \rsp$, $\bTsp$ dominates and the disk annuli tend to be aligned with the planetary spin axis.  For $r \gtrsim \rst$, $\bTst$ dominates and the disk tends to be aligned with the planet's orbit.  For $\rsp \lesssim r \lesssim \rst$, $\bTsg$ dominates and self-gravity strongly influences the disk warp profile.

The equilibrium disk warp profile $\bl(r)$ including the effect of self-gravity is determined by the equation
\be
\label{eq:SSwSG}
\bTst + \bTsp + \bTsg = 0.
\ee
With $\bl(r)$ lying in the plane spanned by $\blp$ and $\bs$, this reduces to
\begin{align}
\label{eq:DimSSwSG}
0 = &z^2 \cos\bg(z) \sin \bg(z)
\nonumber \\
&+ \frac{z_L^5}{z^3} \cos \big[ \bg(z) - \bgp \big] \sin \big[ \bg(z) - \bgp \big]
\nonumber \\
&+ \frac{2-p}{3} \sg \int_{\rin/\rout}^1 \der z' \frac{(z')^{1-p}}{\max(z,z')} \chi b_{3/2}^\one (\chi)
\nonumber \\
&\hspace{15mm}\times\cos \big[ \bg(z) - \bg(z') \big]
\sin \big[ \bg(z) - \bg(z') \big] .
\end{align}

\begin{figure}
\centering
\includegraphics[scale=0.4]{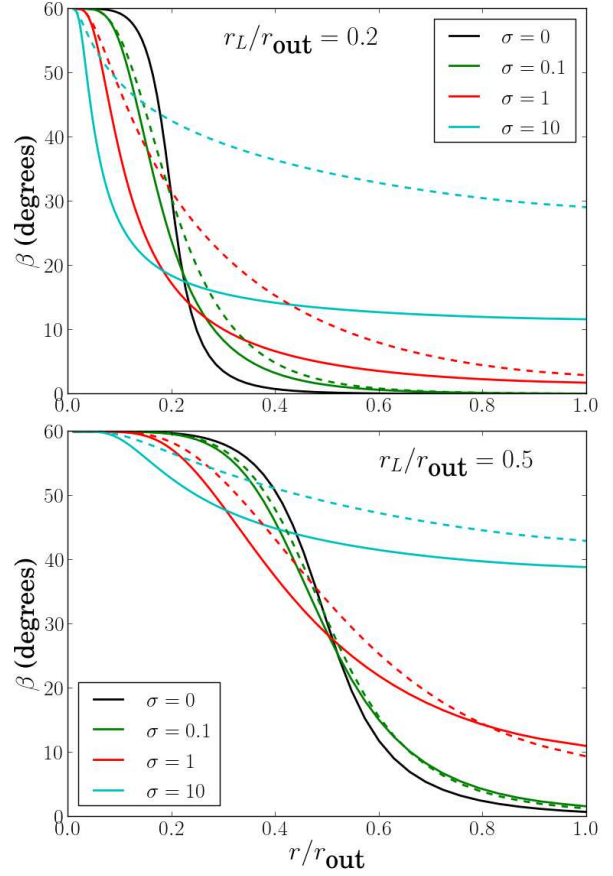}
\caption{Equilibrium disk inclination profile $\bg(r)$ including the effect of self gravity (the generalized Laplace surface), for different values of $r_L/\rout$ and $\sg$ [see Eq.~\eqref{eq:sigma}] as indicated.  The planetary obliquity is assumed to be $\bgp = 60^\circ$.  The $\sg = 0$ curves correspond to the standard Laplace surface (without self-gravity).  The solid lines are for the surface density power-law index $p=1$, and dashed lines for $p=1.5$.}
\label{fig:profiles}
\end{figure}

Figure \ref{fig:profiles} depicts a sample of the equilibrium disk
inclination profile $\bg(r)$ for $r_L/\rout = 0.2, 0.5$ and $p=1,
1.5$, with various values of the disk mass parameter $\sg$.  As
expected, for sufficiently large $\sg$, self-gravity can significantly
increase the outer disk's inclination.  

\begin{figure}
\centering
\includegraphics[scale=0.42]{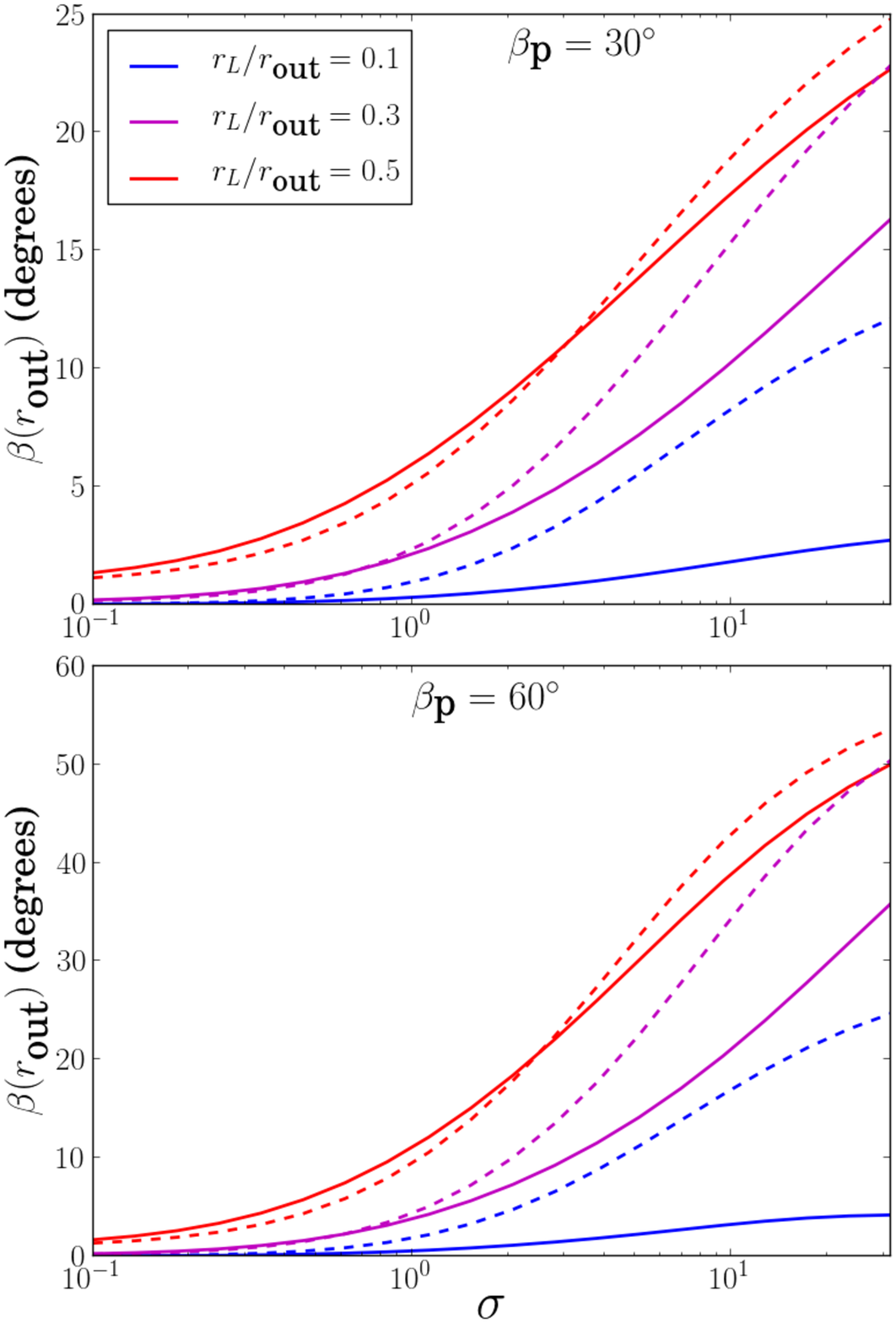}
\caption{Equilibrium inclination of the disk at the outer radius [the angle between $\bl(\rout)$ and $\blp$], as a function of the disk mass parameter $\sg$ [see Eq.\eqref{eq:sigma}].  The top panel is for the planetary obliquity $\bgp = 30^\circ$, and the lower panel for $\bgp = 60^\circ$.  Different colored curves correspond to different values of $r_L/\rout$ as indicated.  The solid lines are for the surface density profile of $p=1$, while the dashed lines are for $p=1.5$.}
\label{fig:betaout}
\end{figure}

Figure \ref{fig:betaout} shows the outer disk inclination angle
$\bg(\rout)$ as a function of $\sg$.  Decreasing the parameter $p$ or
$r_L/\rout$ results in a decrease of $\bg(\rout)$. This can be
understood as follows: The disk inside $r_L$ is roughly aligned with
the planet's spin. This inner disk, together with the planet's
intrinsic quadrupole, act on the outer disk to resist the tidal
torque from the host star and generate $\bg(\rout)$. Reducing $p$
leads to a smaller effective quadrupole of the inner disk, and
therefore yielding a smaller $\bg(\rout)$.

The qualitative behavior of Fig.~\ref{fig:betaout} at low $\sg$
may be understood analytically.  
For $\beta(\rout)\ll \beta_p$, we use the approximate solution 
$\bg(r) \sim \bgp \Theta[(r/\rout)-(r_L/\rout)]$
in the integrand of Eq.~\eqref{eq:DimSSwSG} ($\Theta$ is the Heavyside step function). We find, to an order of magnitude,
\begin{multline}
\label{eq:anbetaout}
\bg(\rout) \sim \left[ \left( \frac{r_L}{\rout} \right)^5 + \left(
  \frac{2-p}{4-p} \right) \sg \left( \frac{r_L}{\rout} \right)^{4-p}
  \right] \\
\times \cos\bgp \sin\bgp. 
\end{multline}
Comparing to Eq.~\eqref{eq:betalap}, the second
term in Eq.~\eqref{eq:anbetaout} may be understood as the correction
to the planet's effective quadrupole due to inner disk's self-gravity.
We see that in order to achieve significant $\bg(\rout)$, both 
$\sg$ and $r_L/\rout$ must be sufficiently large. 
We note that while Eq.~\eqref{eq:anbetaout} captures the correct trend of how
$\beta(\rout)$ depends on $\sg$, $r_L/\rout$ and $p$, it is necessary
to solve Eq.~\eqref{eq:DimSSwSG} to obtain the quantitatively accurate result depicted in Fig.~\ref{fig:betaout}.


\section{Time Evolution of Disk Warp}

In this Section, we first use numerical integrations to examine the
stability property of the generalized Laplace Surfaces obtained in
Section 3. We then consider the possibility of coherent precession of warped self-gravitating disks.

\subsection{Stability of Generalized Laplace Equilibria}

In \cite{Tremaine(2009)}, it was found that the solutions to
Eq.~\eqref{eq:steady} (without disk self-gravity) 
were unstable when $\bgp > 90^\circ$.  Although in this paper
we only consider disk warp profiles with $\bgp < 90^\circ$, it is not
immediately obvious if the addition of self-gravity changes the
stability of the generalized Laplace surfaces obtained by solving 
Eq.~\eqref{eq:SSwSG}. A complete analysis of the Laplace equilibria
[which we denote by $\bl_0(r)$] would require one to find the full
eigenvalue spectrum of the perturbed equation of motion for $\bl(r,t)$.
We do not carry out such an analysis here. Instead, we use numerical integrations
to examine how a small deviation of $\bl(r,t)$ from $\bl_0(r)$ evolves
in time.

The evolution equation for the disk warp profile $\bl(r,t)$ is
\be
\label{eq:TimeEq}
 r^2 \Om \frac{\pd \bl}{\pd t} = \bTst + \bTsp + \bTsg ,
\ee
where $\Om(r) = \sqrt{G \Mp/r^3}$. The small perturbation
$\bj \equiv \bl(r,t) - \bl_0(r)$ satisfies
\be
r^2\Om \frac{\pd \bj}{\pd t} = \bTst + \bTsp + \bTsg.
\ee
We consider two indepedent initial perturbations:
\be
\label{eq:out}
\bj(r,t=0) = 0.02 \sin \left[ \frac{\pi (r - \rin)}{\rout - \rin} \right] \left( \frac{\bs \btimes \blp}{|\bs \btimes \blp|} \right)
\ee
and
\be
\label{eq:in}
\bj(r,t=0) = 0.02 \sin \left[ \frac{\pi (r - \rin)}{\rout - \rin} \right] \left( \frac{\bl_0 \btimes (\bs \btimes \blp)}{|\bl_0 \btimes (\bs \btimes \blp)|} \right).
\ee
Equation \eqref{eq:out} corresponds to a perturbation perpendicular to 
the plane spanned by the Laplace surface, while Eq.~\eqref{eq:in} corresponds to 
a slight change in the disk inclination profile $\bg (r)$. 
We choose the $r$-dependence
in Eqs.~\eqref{eq:out} and \eqref{eq:in} such that $\bj=0$ at $r=\rin$ and
$r=\rout$.

\begin{figure*}
\centering
\includegraphics[scale=0.42]{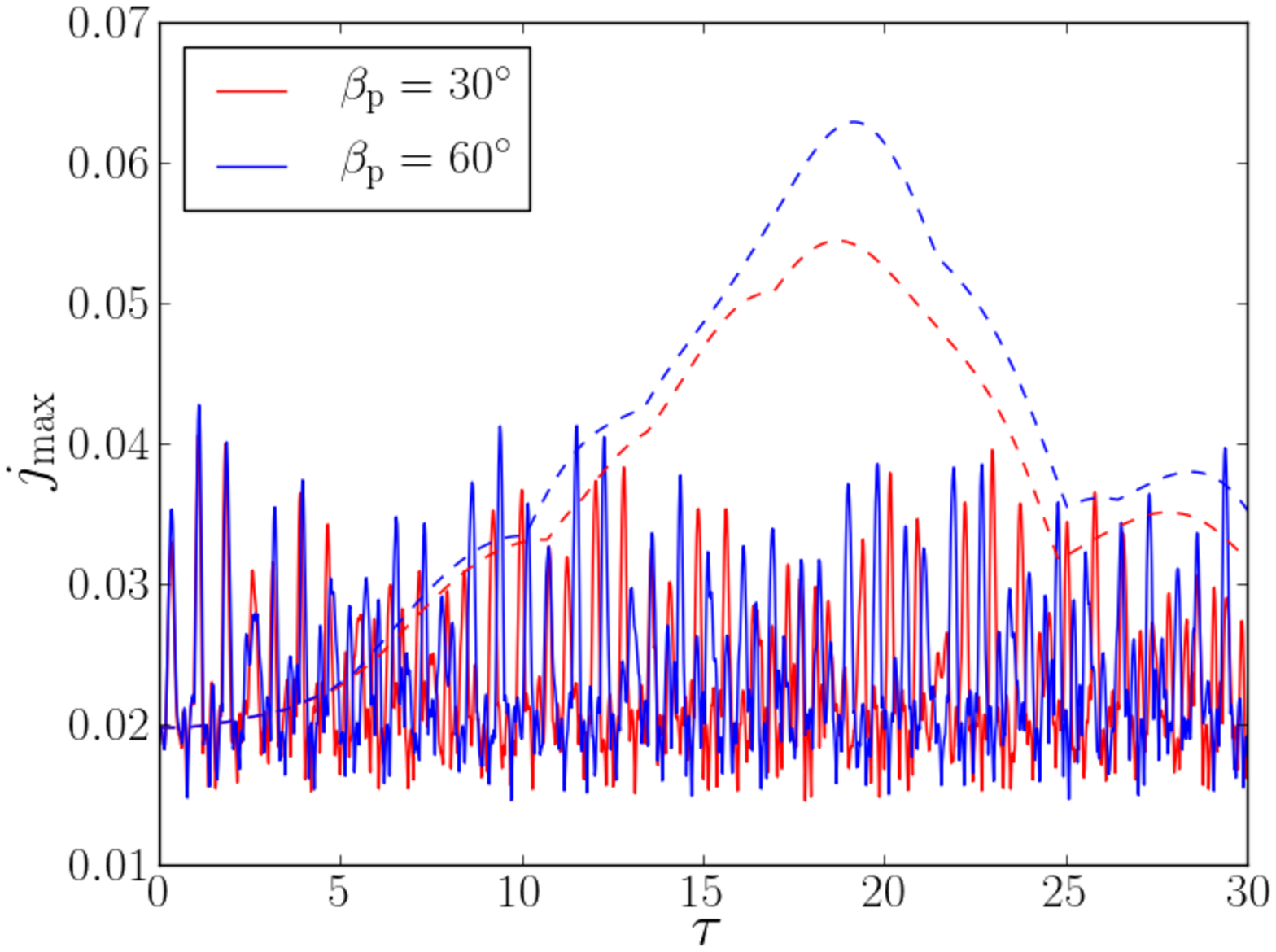}
\includegraphics[scale=0.42]{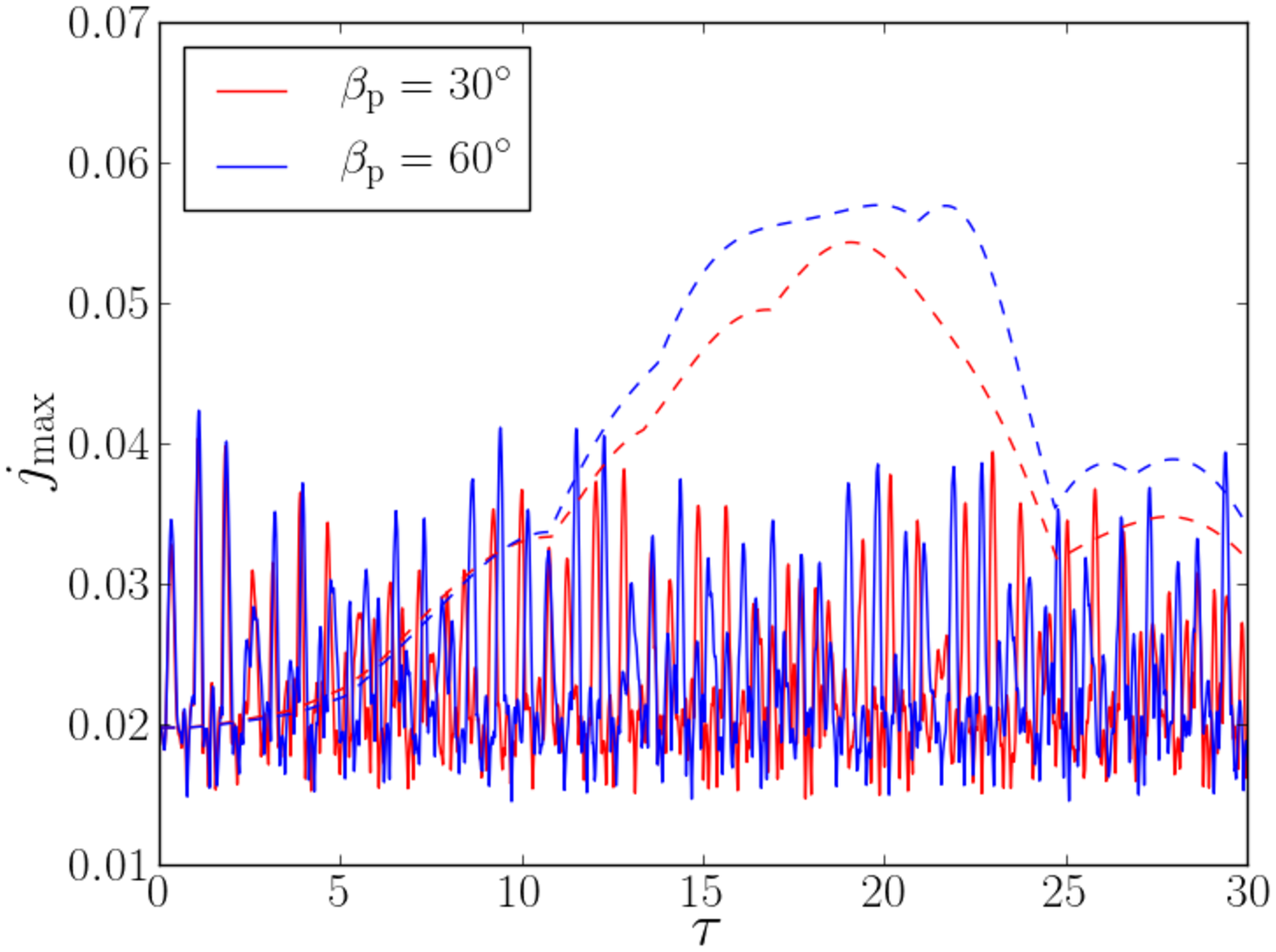}
\caption{Time evolution of the quantity $j_{\rm max}$ [Eq. \eqref{eq:jmax}], with the initial condition given by \eqref{eq:out} (left panel) and \eqref{eq:in} (right panel).    Solid lines denote $\sg = 10$, dotted lines denote $\sg = 0.1$.  Values of $\bgp$ are as indicated.}
\label{fig:jmax}
\end{figure*}

Figure \ref{fig:jmax} shows some examples of our numerical integration results.
We define the quantity
\be
\label{eq:jmax}
j_{\rm max}(t) \equiv \max_{r \in [\rin,\rout]}(|\bj(r,t)|),
\ee
and plot $j_{\rm max}$ for the initial
conditions \eqref{eq:out} and \eqref{eq:in}, with parameters $\bgp =
30^\circ, 60^\circ$ and $\sg = 0.1,10$.  
We see that $j_{\rm max}$ is bounded in all cases. We have carried out 
calculations for other initial conditions and found similar behaviors
for $j_{\rm max}$. We conclude that 
the equilibrium profile $\bl_0(r)$ are stable (for $\bgp<90^\circ$).

In addition to the inclination instability, it was shown in
\cite{Tremaine(2009)} that the Laplace surface (without self-gravity)
is unstable to eccentricity growth when $\bgp \gtrsim 69^\circ$.
This ``eccentricity instability'' cannot be probed by our analysis,
and is beyond the scope of this paper.   All examples considered in this paper have planetary obliquities less than this critical angle.

\subsection{Coherent Disk Precession}
\label{sec:RP}

The generalized Laplace surfaces studied in Section 3 correspond to 
the disk warp equilibria that may be attained when the disk experiences
sufficient internal dissipation. However, we could also imagine 
situations in which circumplanetary disks are formed with a warp profile
that is ``out of equilibrium''.  It is of interest to consider the time
evolution of such ``out-of-equilibrium'' disks. In particular, we are
interested in the following scenario/question: if
a disk is formed with a large inclination at $\rout$ with respect to the
planet's orbit, under what condition can the disk maintain its coherence 
and large inclination at $\rout$?

In general, the disk warp profile $\bl(r,t)$ evolves according to
Eq.~(\ref{eq:TimeEq}). Without self-gravity, the disk will develop
large incoherent warps and twists due to strong differential torques,
and may eventually break. With sufficient
self-gravity, coherent precession of the disk may be possible.

For concreteness, we consider an initially flat disk with $\bl$
aligned with the planet's spin axis $\bs$. Both $\bs$ and $\blp$ are
assumed to be fixed in time, since the planet's spin and orbital angular
momenta are much larger than the disk angular momentum. To determine
the evolution of the disk warp profile, we divide the disk into 30
rings spaced logarithmically in radius, with $r_i$ ($i = 1,2,\dots,30$) ranging from $5\times 10^{-2}r_{\rm out}$ to
$r_{\rm out}$.  We then integrate Eq.~(\ref{eq:TimeEq}) to evolve the
orientation of the individual ring $\bl(r_i,t)$.

\begin{figure*}
\centering
\includegraphics[scale=0.42]{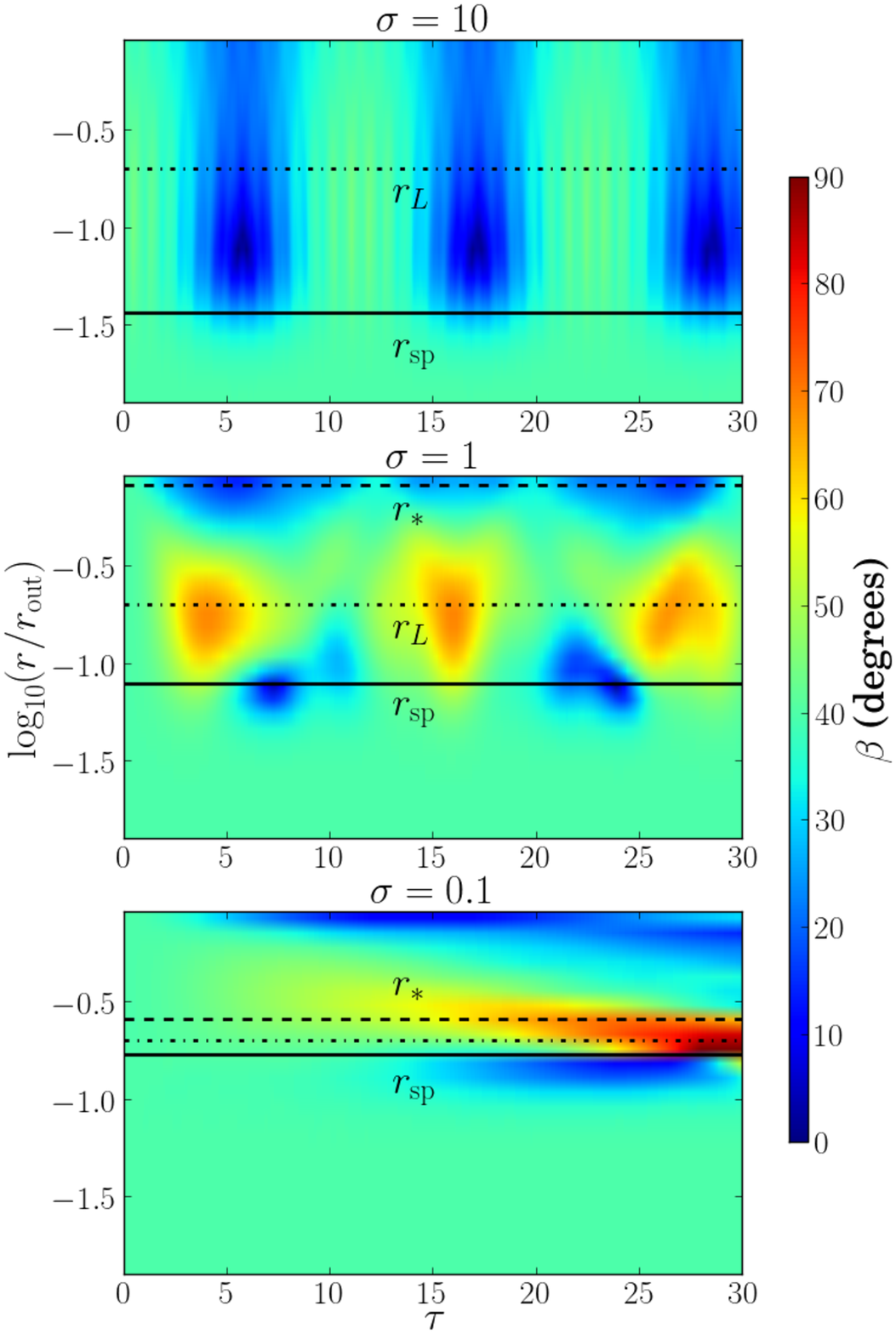}
\includegraphics[scale=0.42]{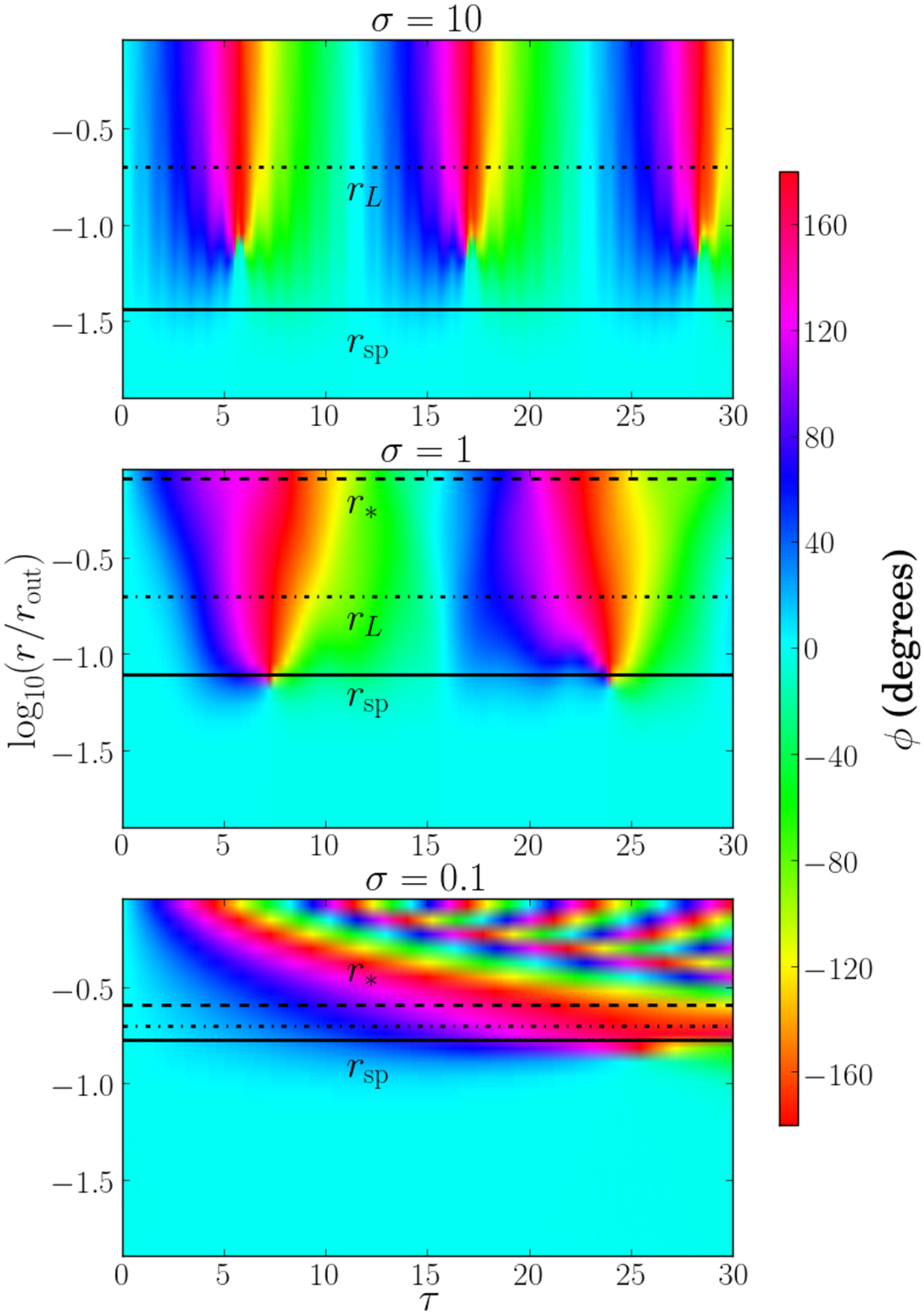}
\caption{Evolution of the disk inclination $\bg(r,t)$ (left panels) and twist angle $\phi(r,t)$ (right panels) for three different disk mass parameters: $\sg = 10$ (top), $\sg = 1$ (middle), and $\sg = 0.1$ (bottom). The dimensionless time is $\tau = t \omst(\rout)$ [see Eq.~\eqref{eq:omst(rout)}].  The horizontal lines mark the locations of $\rsp$ (solid), $\rst/\rout$ (dashed) and $r_L$ (dot-dashed), to indicate where self-gravity and external torques dominate (see Fig.~\ref{fig:torques}).  The planetary obliquity is $\bgp = 40^\circ$ and the Laplace radius is $r_L/\rout = 0.2$.}
\label{fig:angletime}
\end{figure*}

\begin{figure*}
\centering
\includegraphics[scale=0.42]{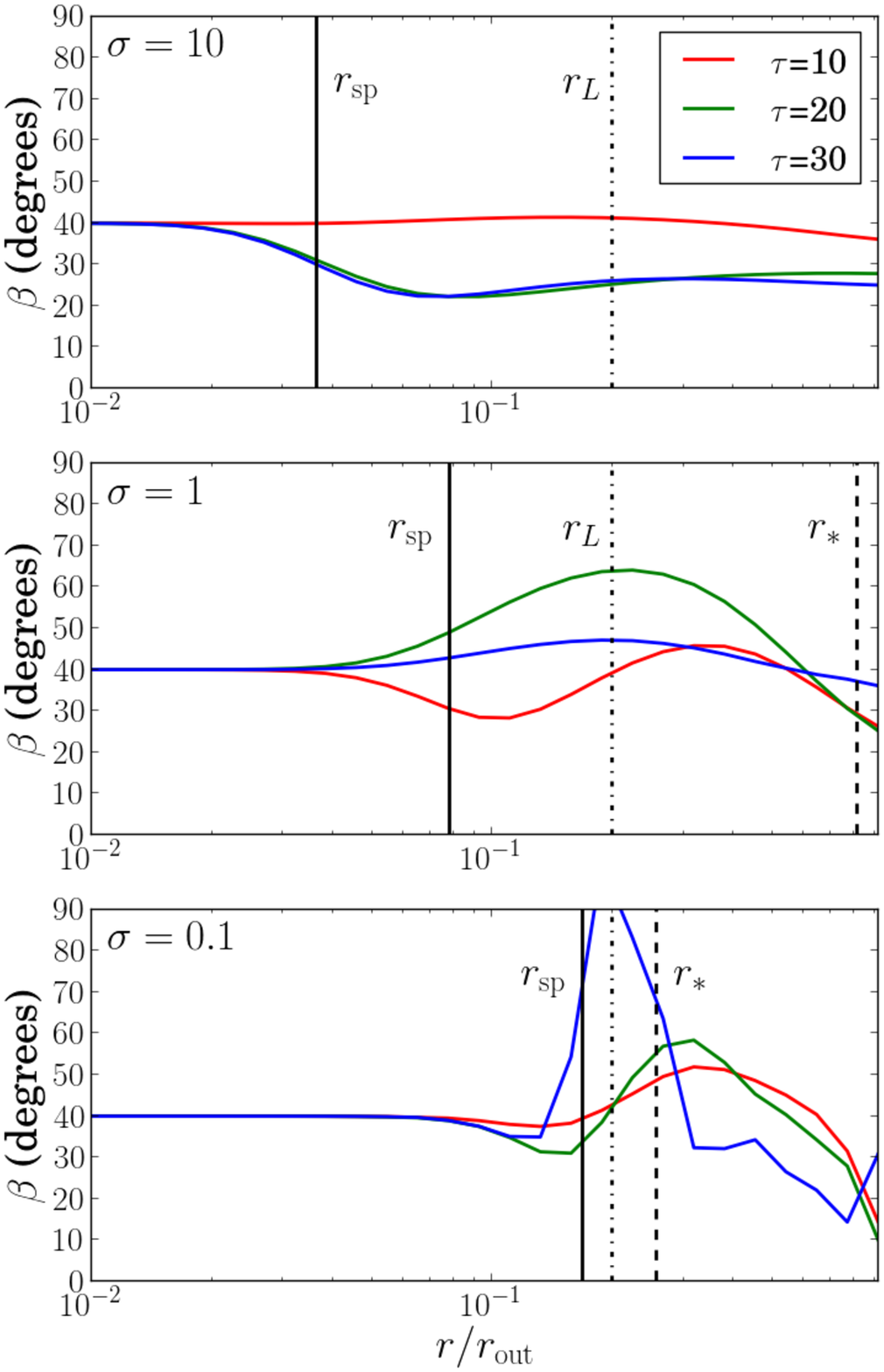}
\includegraphics[scale=0.42]{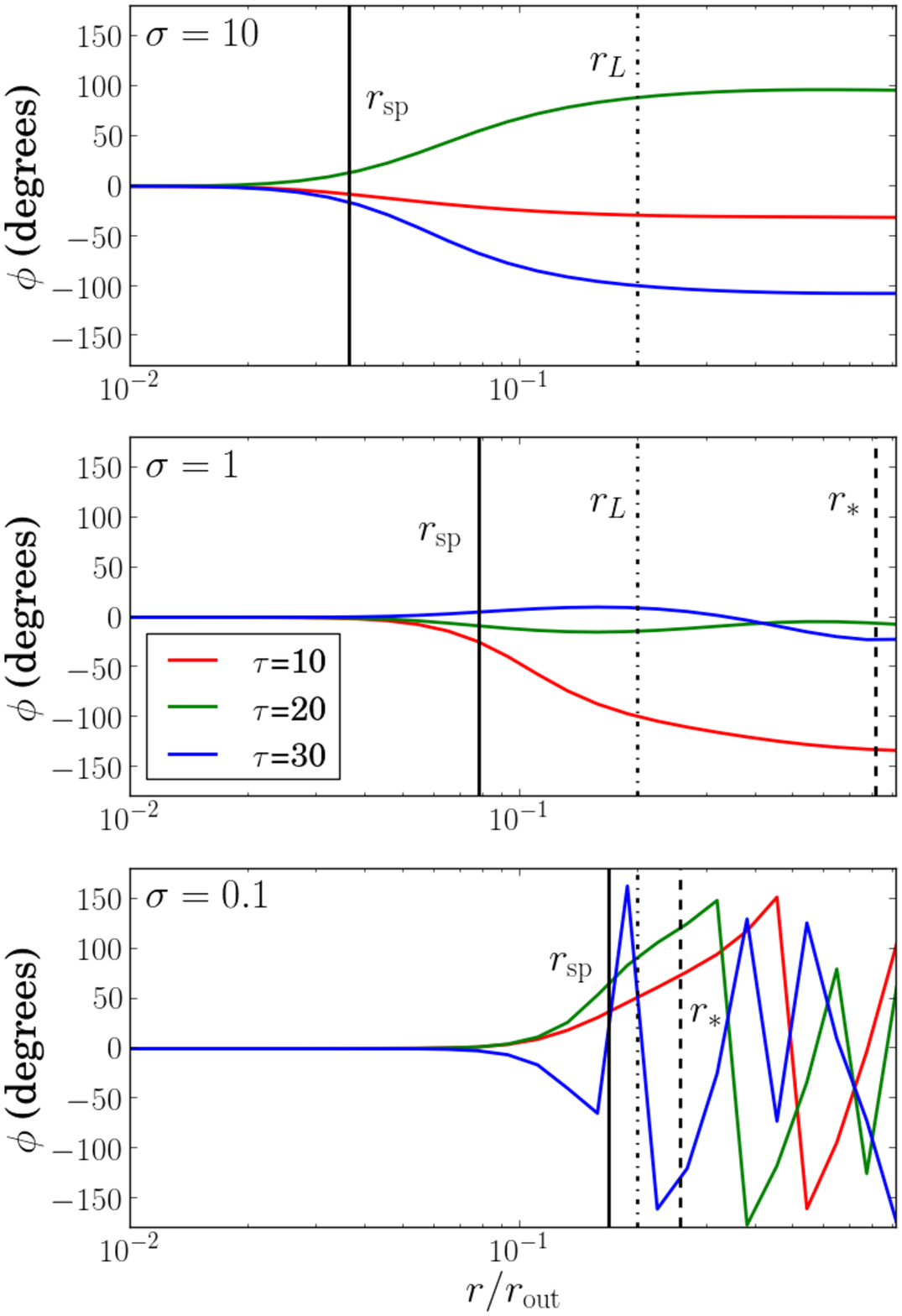}
\caption{Snapshots of the disk inclination profile $\bg(r,t)$ (left) and twist profile $\phi(r,t)$ (right) at $\tau = 10$ (red), $\tau = 20$ (green) and $\tau = 30$ (blue), for the evolution depicted in Fig.~\ref{fig:angletime}.  The vertical lines mark the locations $\rsp$ (solid), $\rst$ (dashed) and $r_L$ (dot-dashed), indicating where self-gravity and external torques dominate.}
\label{fig:betasnap}
\end{figure*}

Figures~\ref{fig:angletime} and \ref{fig:betasnap}
show a sample numerical result, for 
integration time up to $\tau = t \om_*(\rout) =30$, where
\be
\label{eq:omst(rout)}
\omst(\rout) = \frac{3 G \Ms}{4 a^3 \Om(\rout)}
\ee
is the (approximate) precession frequency of the outer disk annulus
torqued by the central star. The planetary obliquity is fixed at
$\bgp = 40^\circ$, with $p=1$ and $r_L/\rout = 0.2$.  We consider
three values of $\sigma$: 10, 1 and 0.1.  In addition to the disk
inclination angle $\beta(r,t)$ [the angle between $\bl(r,t)$ and
  $\blp$], we also show the disk twist angle $\phi(r,t)$ [the angle
  between $\blp\times \bl (r,t)$ and $\blp\times \bs$].  
In all three cases, when $r \lesssim \rsp$ the disk annuli remain mostly aligned
with the planetary spin, with $\bg \approx \bgp = 40^\circ$.  
For the $\sg = 10$ case, the region of the disk beyond
$r_{\rm sp}$ precesses coherently, while for the low-mass case ($\sg =
0.1$), the disk's self-gravity is not able to enforce coherence, 
since different disk annuli precess at different rates.
This transition of the coherent behavior occurs at $r_\star\sim \rout$, or equivalently $\sg \sim 1$. From Eq.~\eqref{eq:rst} we have
\be
{r_\star\over \rout}= \left[ \frac{2(2-p)}{3} \sg \right]^{1/(1+p)}.
\ee
Thus, coherent precession of the outer disk requires $\sg\gtrsim 1$, or in terms of disk mass,
\be
\label{eq:Md>}
\Md \gtrsim 2.67 \times 10^{-3} \Mp \left( \frac{\rout}{0.2 \, r_H} \right)^3.
\ee

When the disk mass is high ($\sg \gg 1$), the dynamical behavior is relatively simple.  This may be understood with the model put forth in the next section.


\subsection{Model for high $\sg$ disk}

We assume that for radii $r < \rsp$, the disk annuli stay aligned with
the oblate planet, while for $r \ge \rsp$ the disk is
a rigid plate being torqued externally by the star and the oblate planet
[see Eqs.~\eqref{eq:Tst} and \eqref{eq:Tsp}].  
In other words, we model the disk inclination profile as
\be
\label{eq:lflat}
\bl(r,t) =
\left\{
\begin{array}{cc}
\bs & r < \rsp \\
\bn(t) & r \ge \rsp
\end{array}
\right.  ,
\ee
with $\bn$ evolving in time according to
\be
\label{eq:ntimeev}
\frac{\der \bn}{\der t} = \pomst (\bn \bcdot \blp)(\blp \btimes \bn) + (\pomsp + \bar \omega_\text{d,in})
 (\bn \bcdot \bs)(\bs \btimes \bn),
\ee
where
\begin{align}
\label{eq:pomst}
\pomst &= \frac{2\pi}{L_\text{d,out}} \int_{\rsp}^\rout \Sg(r)r \left(\frac{3 G \Ms r^2}{4 a^3} \right) \der r, \\
\label{eq:pomsp}
\pomsp &= \frac{2\pi}{L_\text{d,out}} \int_{\rsp}^\rout \Sg(r) r \left( \frac{3 G \Mp \Rp^2 J_2}{2 r^3} \right) \der r , \\
\label{eq:pomdin}
\bar \omega_\text{d,in} &= \frac{2\pi}{L_\text{d,out}} \int_{\rsp}^\rout \Sg(r) r \left(\int_{\rin}^{\rsp} \frac{3 \pi G \Sg(r') (r')^3}{2 r^3} \der r' \right) \der r, \\
\label{eq:Ldout}
L_\text{d,out} &= 2\pi \int_{\rsp}^\rout \Sg(r) r^3 \Om(r) \der r.
\end{align}
and $\omst(\rout)$ is given by Eq.~\eqref{eq:omst(rout)}.  Note that $r_{\rm sp}$ depends on $\sg$ [see Eq.~\eqref{eq:rsp} and
Fig.~\ref{fig:torques}].   Assuming $\rin \ll \rsp \ll \rout$, 
\begin{align}
\label{eq:pomstapprox}
\pomst &\simeq \omst(\rout) \frac{5-2p}{2(4-p)}, \\
\label{eq:pomspapprox}
\pomsp &\simeq \omst(\rout) \frac{5-2p}{2(1+p)} \left( \frac{r_L}{\rout} \right)^5 \left( \frac{\rout}{\rsp} \right)^{1+p}, \\
\label{eq:pomdinapprox}
\bar \omega_\text{d,in} &\simeq \omst(\rout) \frac{(5-2p)(2-p)}{2(4-p)(1+p)} \sg \left( \frac{\rsp}{\rout} \right)^{3-2p}.
\end{align}

\begin{figure}
\centering
\includegraphics[scale=0.4]{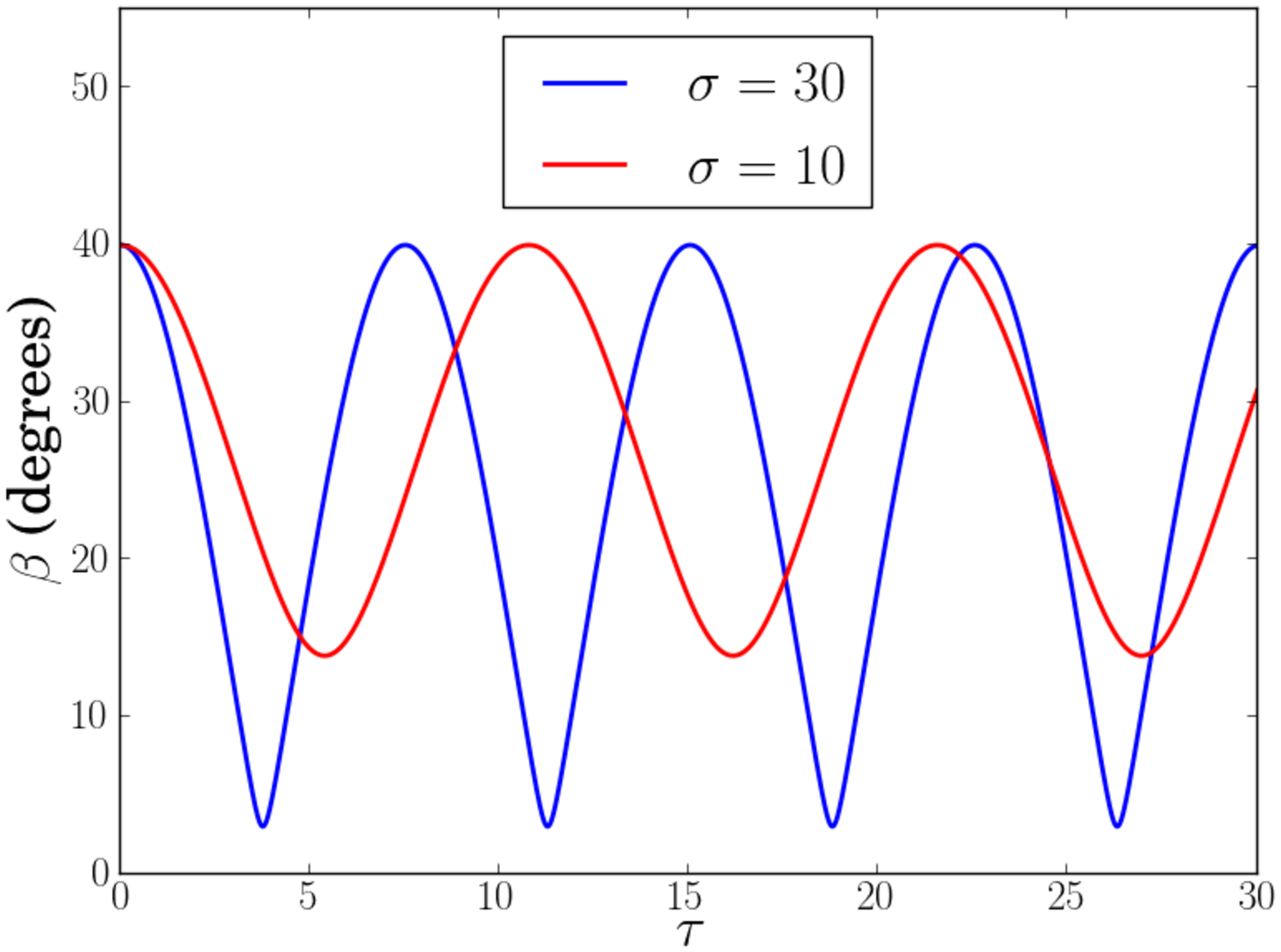}
\includegraphics[scale=0.4]{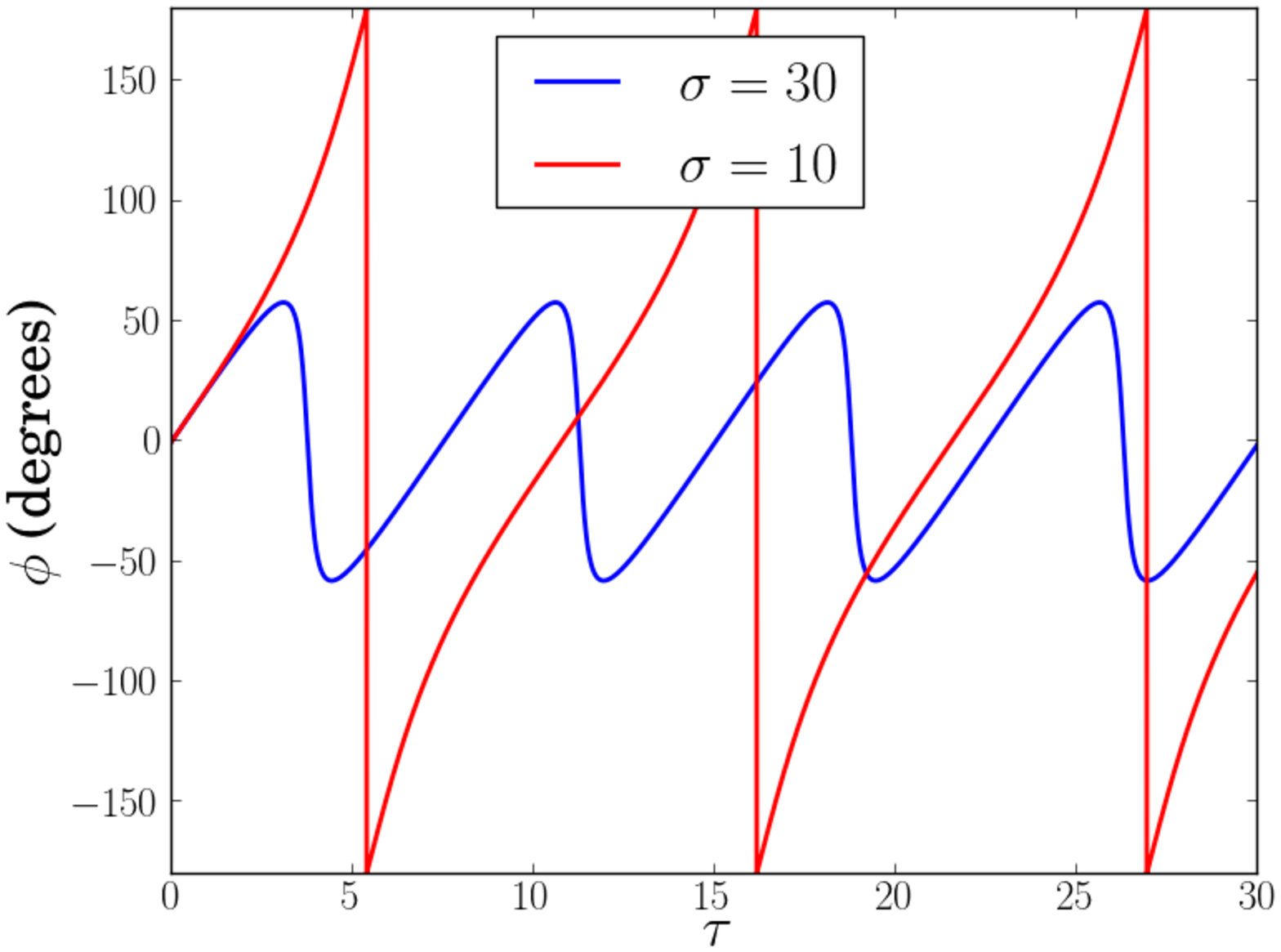}
\caption{Evolution of the (flat) outer disk inclination $\bg$ and twist angle $\phi$ for the simple model [see Eqs. \eqref{eq:lflat}-\eqref{eq:ntimeev}], with two values for the disk mass parameter $\sg$ as indicated.  For $\sg = 30$, the disk normal vector $\bn$ precesses around the planetary spin vector $\bs$, with $\phi$ librating around $\phi = 0^\circ$, while $\bg$ varies from $0^\circ$ to $40^\circ$.  For $\sg = 10$, the outer disk precesses fully around the planetary orbital angular momentum axis $\blp$, indicated by $\phi$ spanning the full range of $-180^\circ$ to $180^\circ$, while $\bg$ remains more or less constant.  The Laplace radius is $r_L/\rout = 0.2$, with $p=1$.}
\label{fig:ToyModel}
\end{figure}

In Fig.~\ref{fig:ToyModel}, 
we show the outer disk inclination $\bg$ and precession angle $\phi$ for $\bn$, with $r_L/\rout = 0.2$ and $p=1$.  
The qualitative behavior seen in
Figs.~\ref{fig:angletime} and \ref{fig:betasnap} is reproduced.
In particular, for $\sg=10$, the outer disk undergoes full precession in $\phi$
while the inclination $\beta$ nutates;
for $\sg=30$, the disk librates in $\phi$ around $0^\circ$, with 
$\beta$ varying between $0^\circ$ and $40^\circ$.

In our model, the behavior of $\phi$ switches from precession to libration at
$\sg \approx 23$.


\section{Summary and Discussion}
\label{sec:conc}

\subsection{Key Results}

Motivated by the recent (tentative) observational evidence for the circumplanetary disk/ring system around the young K5 star 1 SWASP J140747-354542 \citep{Mamajek(2012),vanWerkhoven(2014),Kenworthy(2015),KenworthyMamajek(2015)},
we have presented a
general theoretical study of the inclination (warp) profile of extended disks around giant planets (or brown dwarfs).  Such a disk experiences torques from the host star and the oblate planet. In the absence of any internal torque,
the disk may assume an equilibrium warp profile (the Laplace surface; see Section~\ref{sec:System+Torques}), such that the outer disk beyond the
Laplace radius $r_L$ [see Eq.~\eqref{eq:rL}] tends to be aligned with the planet's orbit (see Fig.~\ref{fig:laplace}).
We have studied how self-gravity of the disk affects the steady-state disk inclination profile (Fig.~\ref{fig:profiles}).  In general, for a given planetary obliquity $\bgp$, the outer disk inclination can be increased due to the ``rigidity" provided by the disk's self-gravity.  To produce a non-negligible outer disk misalignment requires that the combination of the disk mass and $r_L/\rout$ be sufficiently large [see Fig.~\ref{fig:betaout} and Eq.~\eqref{eq:anbetaout}]. The required disk mass is larger for smaller $r_L/\rout$. (Of course, if the disk lies completely inside $r_L$, i.e. $r_L/\rout \gtrsim 1$, self-gravity is not needed to achieve misalignment of the disk since $\bg\simeq \bgp$.)

We have shown that the generalized Laplace equilibria for disk warp profiles 
are stable against small inclination perturbations (Section 4.1).
Because a circumplanetary disk may not relax to a steady state in the
absence of internal dissipation, we have also studied the dynamical
evolution of a disk initially aligned with the planet's spin
(Section~\ref{sec:RP}).  Such a disk can attain misalignment with
respect to the orbital plane if it can precess coherently and if
$\bgp \neq 0$.  We showed that to achieve coherent disk precession,
the disk's self-gravity must dominate over the influence of the star's
tidal torque throughout the disk.  This coherence requirement leads to
a lower bound on the disk mass [Eq.~\eqref{eq:Md>}]:
\be
\nonumber
\Md \gtrsim 2.67 \times 10^{-3} \Mp \left( \frac{\rout}{0.2 \, r_H} \right)^3.
\ee 
Of course, this mass constraint is needed only if $\rout > r_L$.


\subsection{Hydrodynamical Effects}

In this paper we have focused on the effect of self-gravity in
maintaining the coherence and inclination of circumplanetary disks.
Here we briefly comment on hydrodynamical effects internal to the disk.

As noted in Section 1, hydrodynamic forces work to keep the
disk coherent through either bending waves or viscosity.
If the disk viscosity parameter $\alpha$ satisfies $\ag \lesssim H/r$,
the warp disturbances propagate through the circumplanetary disk in 
the form of bending waves.  In order to enforce coherence, a bending
wave must propagate throughout the disk faster than a precession
period from the tidal torque of the host star
\citep{Larwood(1996)}.  The tidal precession
period is of order $t_* \sim 2\pi r^2\Omega/|\bTst|\sim (8\pi/\Omega)(r_H/r)^3$,
while the bending-wave crossing time is 
$t_{\rm bend}\simeq  2 r/c_{\rm s}\simeq (2/\Omega) (r/H)$
($c_{\rm s}$ is the disk sound speed). Thus the small value of $H/r$ 
($\sim 10^{-3}$ for the inferred ring system around J1407b) makes
$t_*$ smaller than $t_{\rm bend}$ when the disk extends to a significant
fraction of the Hill radius.

If the disk viscosity parameter satisfies $\ag \gtrsim H/r$, hydrodynamical forces
communicate through the disk in the form of viscosity.  The 
the internal viscous torque (per unit mass) is 
\citep{PapaloizouPringle(1983)}
\be
|{\bm T}_{\rm visc}| = \frac{r^2 \Om^2}{2} \left( \frac{H}{r} \right)
  \left( 3 \ag + \frac{1}{2 \ag} \left| \frac{\pd \bl}{\pd \ln r} \right| \right)
\ee
Comparing this with the tidal torque $|\bTst|$ shows that unless
the disk warp $|\pd\bl/\pd\ln r|$ is significant, 
the viscous torque will have difficulty balancing the tidal torque from the
host star; such a strongly warped disk could be subjected to 
breaking \citep{Dougan(2015)}.

In addition to the above considerations, the ``observed'' gaps in the 
J1407b disk may halt the propagation of bending waves and cut
off viscous torques. Thus, hydrodynamical effects cannot be
responsible for the disk's coherence and inclination.

\subsection{Implications}

The large disk mass [Eq.~\eqref{eq:Md>}] required to enforce coherent disk precession or maintain 
misalignment of the outer disk  may be difficult to achieve in the context of circumplanetary 
disk formation (e.g. \citealt{CanupWard(2006)}). Moreover, a massive disk can suffer gravitational 
instability. Evaluating the Toomre $Q$ parameter at the outer
radius of the disk, we find
\begin{align}
\label{eq:Q}
Q(\rout) &= \frac{c_\text{s}(\rout) \kappa(\rout)}{\pi G \Sg(\rout)} 
\nonumber\\
&\simeq\frac{2}{2-p} \left(\frac{H(\rout)}{10^{-3} \rout} \right) \left( \frac{10^{-3} \Mp}{\Md} \right).
\end{align}
where we have used $c_\text{s} \simeq H \Om$, $\kappa \simeq \Om
\simeq \sqrt{G \Mp/r^3}$ ($H$ is the disk scale-height).  
Requiring $Q\gtrsim 1$ for stability puts an upper limit on $\Md$, and thus
the size of the disk. Combining Eqs. \eqref{eq:Q} and \eqref{eq:Md>}, we find 
\be
\frac{\rout}{r_H}\lesssim 0.35 \left( \frac{H}{10^{-3} \rout} \right)^{1/3}.  
\ee
This puts a strong constraint on the putative ring/disk system around J1407b.


Our work shows that in general, an extended circumplanetary disk is
warped when in a steady state or undergoing coherent precession. This warp depends
on the Laplace radius [see Eq.~\eqref{eq:rL}] and the disk mass.
Direct observations of such a warped circumplanetary disk would
constrain the planet's oblateness (the $J_2$ parameter), complementing photometric constraints \citep{CarterWinn(2010), Zhu(2014)}.



Although our work is motivated by the the putative J1407b ring system,
our results can be easily adapted to circumplanetary disk/ring systems
in general.  We expect that the analysis developed in this paper can
be a useful tool to evaluate the stability of
circumplanetary disk/ring systems detected in the future.

\section*{Acknowledgments}

We thank the referee for providing thorough and thoughtful comments,
which have significantly improved our paper.
This work has been supported in part by NSF
grant AST-1211061, and NASA grants NNX14AG94G and NNX14AP31G.
JZ is supported by a NASA Earth and Space Sciences
Fellowship in Astrophysics.

\section*{Appendix: Exact self-gravity torque for a circular disk}

As noted in Section \ref{sec:SS+SG}, Eq. \eqref{eq:Tsg} is valid only
when $|\bl(r') \btimes \bl(r)| \ll 1$ or $\chi \ll 1$.  When
$\chi \sim 1$ and $|\bl(r') \btimes \bl(r)| \sim 1$, a
different formalism is needed to compute the torque acting between two
circular massive rings.  In terms of the warp profile $\bl(r,t)$ and
disk surface density $\Sg(r)$, the specific torque acting on a disk
annulus at radius $r$ 
from the disk's self-gravity is
\citep{Kuijken(1991),ArnaboldiSparke(1994), Ulubay-Siddiki(2009)}
\begin{align}
\bTsg = \int_{\rin}^{\rout} \der r' & \frac{4\pi G \Sg(r')}{\max(r,r')} \frac{\chi I(\chi,\sin^2\ag)}{(1+\chi^2)^{3/2}} 
\nonumber \\
&\times \big[ \bl(r,t)\bcdot \bl(r',t) \big] \big[ \bl(r,t) \btimes \bl(r',t) \big]
\label{eq:Tsgexact}
\end{align}
where $\chi = \min(r,r')/\max(r,r')$, $\sin^2 \alpha = |\bl(r,t) \btimes \bl(r',t)|^2$, 
\begin{align}
I = &\frac{4}{\pi^2} \int_0^{\pi/2} \der \psi \left[ \frac{E(k)(1-k^2/2)}{(1-k^2)} - K(k) \right]
\nonumber \\
&\times \frac{(1-k^2/2)^{3/2}}{k^2} \frac{\sin^2 \psi }{\sqrt{1-\sin^2 \ag \sin^2 \psi}} \\
k^2 = &k^2(\chi,\sin^2\psi,\sin^2\ag)
\nonumber\\
= & \frac{4 \chi \sqrt{1-\sin^2\ag \sin^2 \psi}}{1 + \chi^2 + 2\chi \sqrt{1-\sin^2\ag \sin^2 \psi} }
\end{align}
while $K(k)$ and $E(k)$ are elliptic integrals of the first and second
kind, respectively.  The only approximation used in the derivation of
Eq.~\eqref{eq:Tsgexact}
is that the disk is infinitesimally thin; this formula is
exact for arbitrary $\chi$ and mutual inclination angles
$\alpha$.

\begin{figure}
\centering
\includegraphics[scale=0.4]{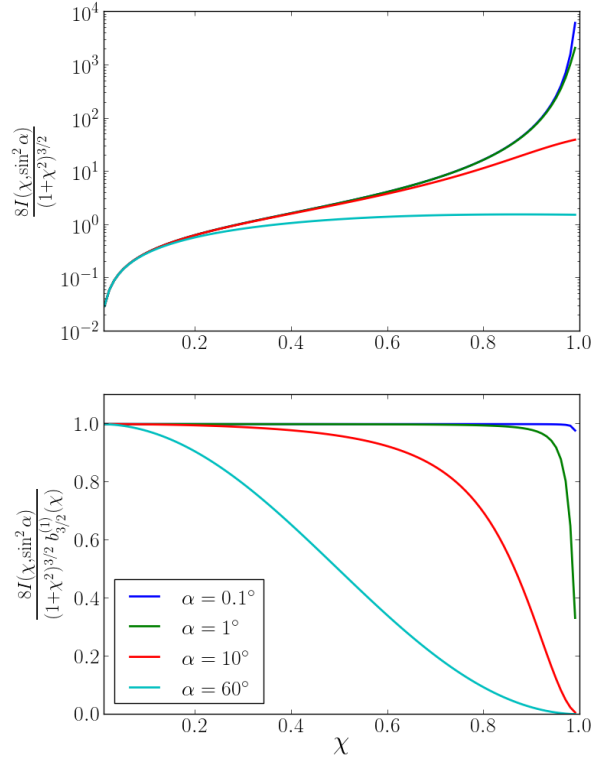}
\caption{The top panel shows the integrand in Eq. \eqref{eq:Tsgexact} as a function of $\chi$, with values of $\ag$ as indicated.  We remove the dependence on $\sin\ag \cos\ag$.  The bottom plot shows the ratio of the integrand in \eqref{eq:Tsgexact} and
that in \eqref{eq:Tsg}.}
\label{fig:intchi}
\end{figure}

In the top panel of Fig. \ref{fig:intchi}, we plot the integrand in
equation \eqref{eq:Tsgexact},
\be
\label{eq:intchi}
\frac{8 I(\chi, \sin^2 \ag)}{(1+\chi^2)^{3/2}},
\ee
as a function of $\chi$. We remove the dependence of $\sin\ag \cos\ag$, as 
they are already present in our approximation \eqref{eq:Tsg}. 
We see that when $|\ag| > 0$, the integrand \eqref{eq:intchi} becomes
large but stays finite as $\chi \to 1$.  In the bottom panel of
Fig.~\ref{fig:intchi}, we plot the ratio of the integrands in
Eqs.~\eqref{eq:Tsgexact} and \eqref{eq:Tsg},
\be
\label{eq:ratio}
\frac{8 I(\chi, \sin^2 \ag)}{(1+\chi^2)^{3/2} \; b_{3/2}^{(1)}(\chi)}.
\ee
Since the quantity \eqref{eq:ratio} is approximately unity for most of
the parameter range of interest (Fig. \ref{fig:intchi}), we do not
expect significant corrections to the equilibrium disk warp profiles
obtained in Section \ref{sec:SS+SG}.

\begin{figure}
\centering
\includegraphics[scale=0.4]{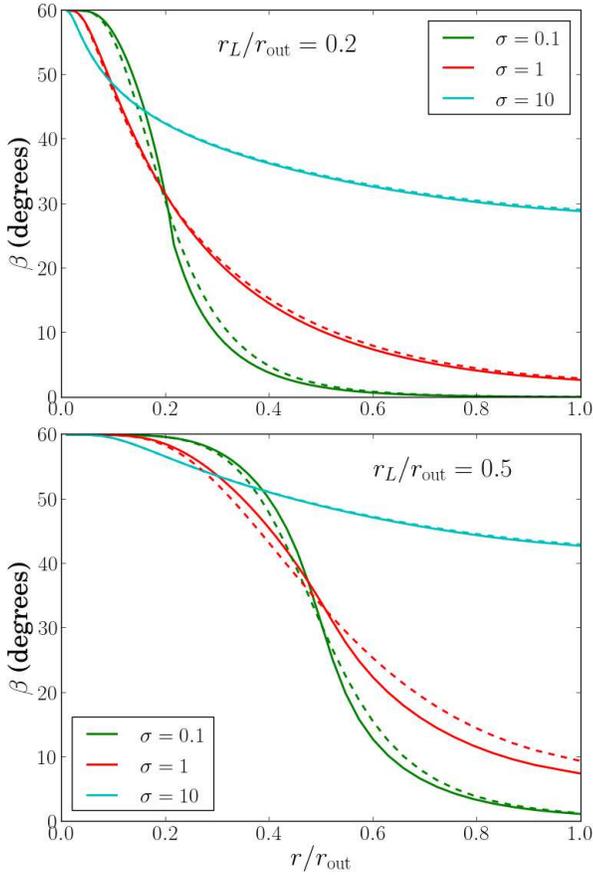}
\caption{Equilibrium disk inclination profile $\bg(r)$
  including the effect of self-gravity for different values of
  $r_L/\rout$ and $\sigma$ [see Eq. \eqref{eq:sigma}] as indicated.
  The results obtained using the approximate self-gravity torque 
[Eq. \eqref{eq:Tsg}] are shown in dashed lines, while those obtained with 
the exact self-gravity torque [Eq. \eqref{eq:Tsgexact}] are shown by solid lines.  
We take $p=1.5$ [Eq. \eqref{eq:Sg}] and $\bgp = 60^\circ$ for all
  solutions.}
\label{fig:exactsols}
\end{figure}

We have repeated the calculation of the Laplace equilibria for disk warp profiles
using the exact torque expression \eqref{eq:Tsgexact}. 
Figure \ref{fig:exactsols} shows a sample of our numerical results
for the disk inclination profile $\bg(r)$, with $\Sigma \propto
r^{-3/2}$ and the values of $\sg$ and $r_L/\rout$ as indicated.  
The solutions for $\bg(r)$ with the approximate torque expression
\eqref{eq:Tsg} are also shown for comparison.
We see that using the exact self-gravity
torque \eqref{eq:Tsgexact} changes the solution of the equilibrium
disk warp $\bg(r)$ by less than a few degrees in all cases.

\end{document}